\documentclass[12pt]{article}
\usepackage{graphicx}

\newcommand\pubnumber{}
\newcommand\pubdate{\today}

\def\institute{Department of Physics and Astronomy\\
University of Notre Dame, Notre Dame, IN 46556, USA}

\def\Title#1{\begin{center} {\Large #1 } \end{center}}
\def\Author#1{\begin{center}{ \sc #1} \end{center}}
\def\Address#1{\begin{center}{ \it #1} \end{center}}

\newcommand\pubblock{\rightline{\begin{tabular}{l} \pubnumber\\
         \pubdate  \end{tabular}}}
\newenvironment{Abstract}{\begin{quotation}  }{\end{quotation}}
\newenvironment{Presented}{\begin{quotation} \begin{center} 
             PRESENTED AT\end{center}\bigskip 
      \begin{center}\begin{large}}{\end{large}\end{center} \end{quotation}}

\newcommand{\tth}{\mathrm{t\bar{t}H}}
\newcommand{\ttv}{\mathrm{t\bar{t}V}}

\newcommand{\ttbar}{\mathrm{t\bar{t}}}

\begin{document}
\begin{titlepage}
\pubblock

\vfill
\Title{2016 Update of the $\tth$ Multilepton Analysis at $\sqrt{s} = 13$ TeV}
\vfill
\Author{Charles Mueller}
\Address{\institute}
\vfill
\begin{Abstract}
The latest results from the search for a Standard Model Higgs boson produced in
association with a top quark pair ($\tth$) at $\sqrt{s} = 13$ TeV decaying to final states with
multiple leptons is presented using the 2016 dataset from the CMS experiment.
The Higgs decays into either WW*, ZZ*, or $\tau\tau$, and the top quark pair decays
considered are either fully leptonic, or semi-leptonic. The leptons defining the
final states are muons and/or electrons. The overall analysis strategy, as well
as new techniques with respect to the 2015 results are outlined.
\end{Abstract}
\vfill
\begin{Presented}
$9^{th}$ International Workshop on Top Quark Physics\\
Olomouc, Czech Republic,  September 19--23, 2016
\end{Presented}
\vfill
\end{titlepage}
\def\thefootnote{\fnsymbol{footnote}}
\setcounter{footnote}{0}

\section{Introduction}
While the discoveries of the top quark and Higgs boson were essential in
verifying the standard model (SM) of particle physics, many important questions
remain unanswered. Specifically, why SM particles have the masses that are observed,
and whether or not the top quark's large mass comes only from it's interaction with the Higgs.
The top quark is the heaviest fundamental particle, suggesting
a special role in electroweak symmetry breaking. The top quark's large mass comes
from its Yukawa interaction with the Higgs. This property of the top quark
makes $\tth$ processes an excellent place to test the SM, while remaining
sensitive to new physics beyond the SM. By comparing the $\tth$ coupling with SM Higgs
production via a gluon-gluon loop with virtual top exchange, limits can be set on
new physics in the gluon-gluon loop. 

The analysis presented here covers the latest 2016 LHC Run II $\tth$ result~\cite{cms-hig-16-022},
measured by the CMS detector~\cite{cms-jinst}. The Higgs decays targeted include
WW*, ZZ*, and $\tau\tau$. The Higgs and top systems may decay semi-leptonically or fully
leptonically, but the final state leptons must originate from both the top and Higgs systems.
The experimental signatures include two same-sign leptons and $\ge$three leptons, where leptons are defined as
muons or electrons. 
Placing these requirements greatly narrows the available phase space, making
signal processes rare, but eliminates the largest backgrounds.  

\section{Object selection}
The most important aspect of the physics objects in this analysis is the lepton selection.
The lepton selection uses 3 distinct classes of increasingly tight categories. The first category
of leptons is the preselection, consisting of cuts on kinematics, isolation, vertexing and experiment-specific identification
variables. The next class is the fakeable selection, which is enriched in non-prompt leptons originating from b-quark decays.
These leptons are used later for defining control regions. The final lepton category, known as the ``tight'' selection is used to select
prompt leptons for the signal region definitions. This selection includes tightened versions of the preselection cuts, and
additionally a cut on a multivariate analysis (MVA) value used specifically to identify prompt from non-prompt leptons. The discriminating
power of this lepton MVA is shown below in Figure~\ref{fig:lep_mva}. This lepton
MVA uses lepton variables related to isolation, vertexing, nearest jet, and detector-specific identification as inputs to discriminate prompt
from non-prompt leptons.

\begin{figure}[htb]
\centering
\includegraphics[height=2in]{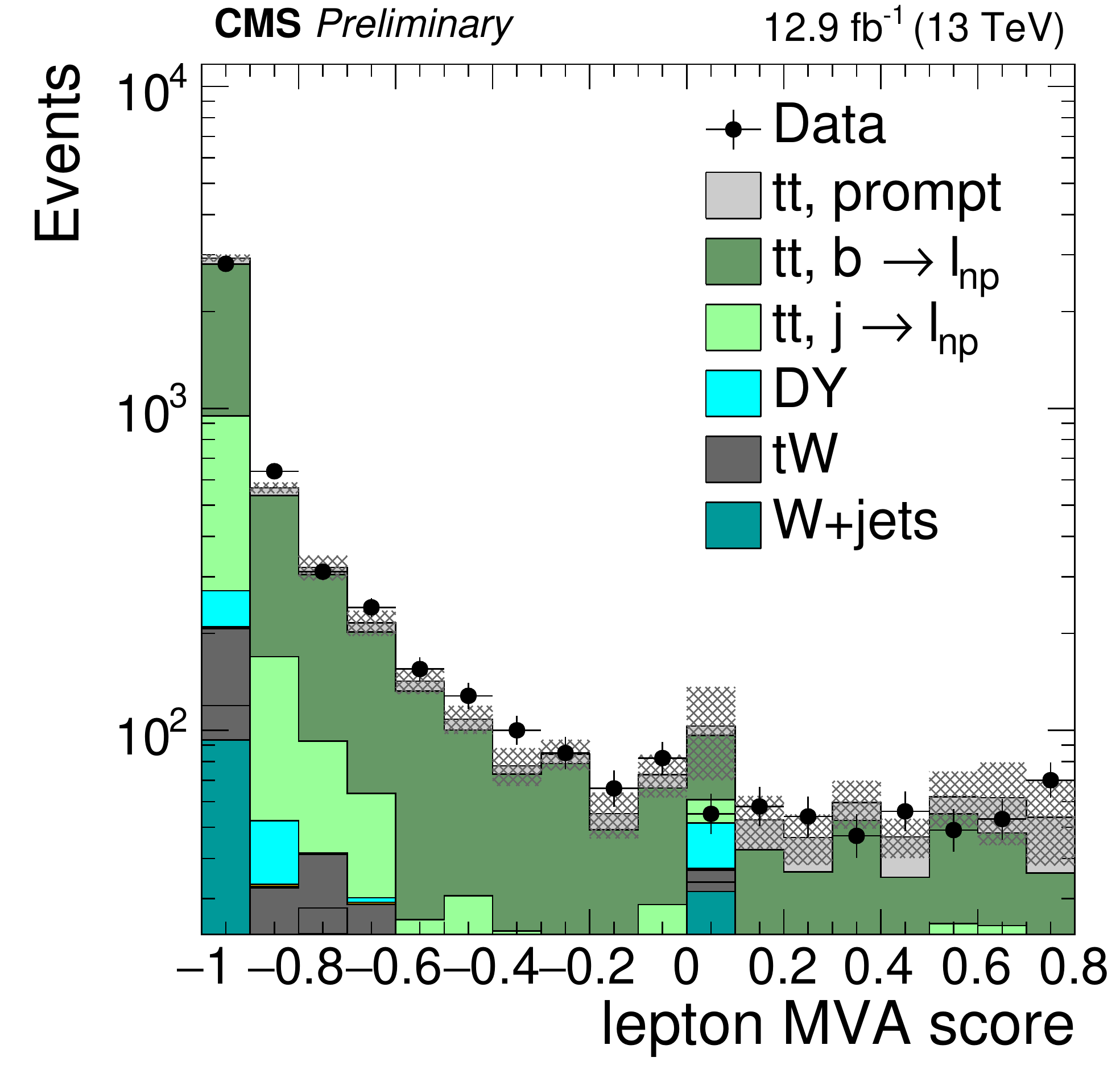}
\includegraphics[height=2in]{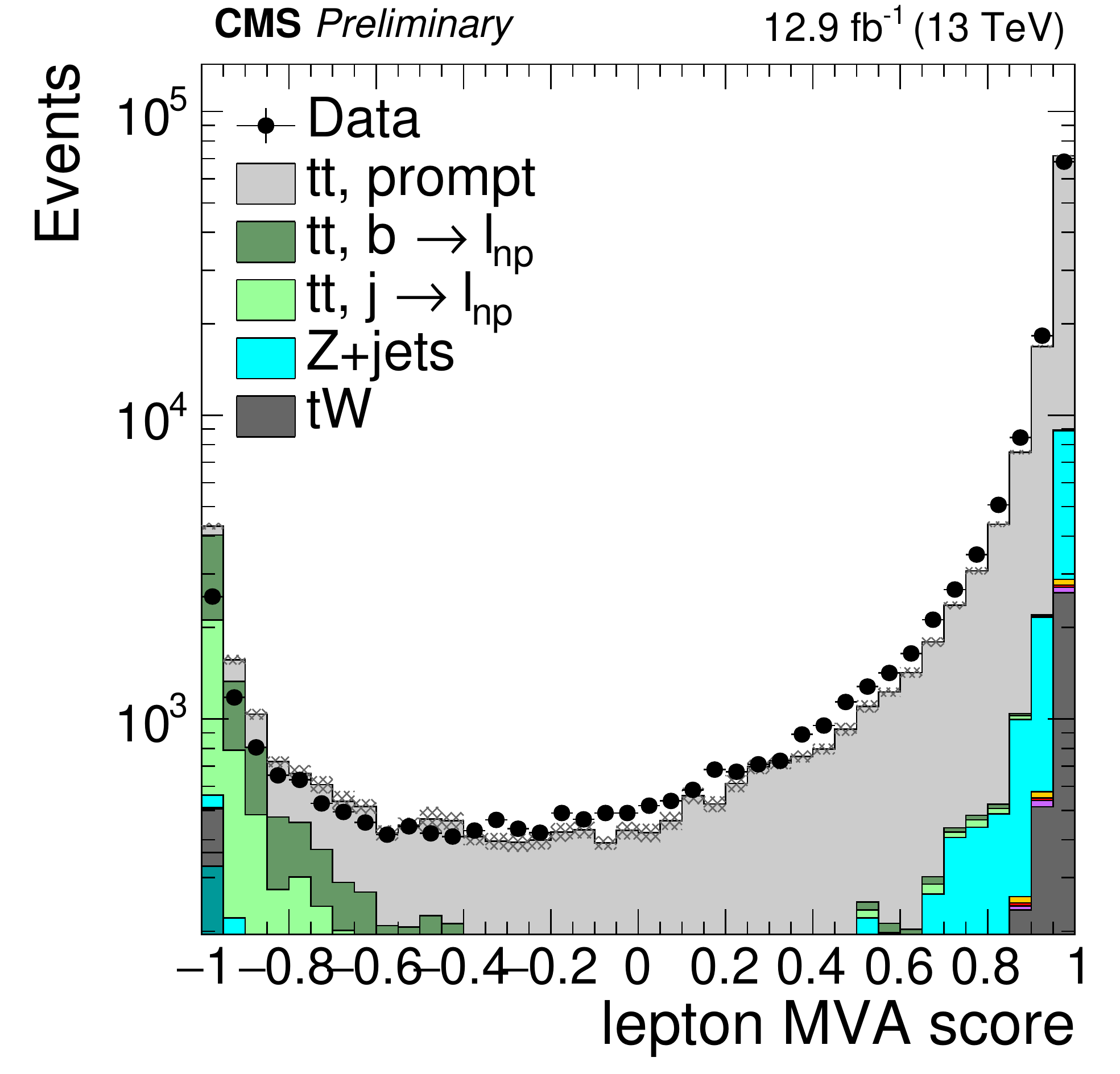}
\caption{The prompt lepton MVA output in control regions enriched in non-prompt leptons (left) and enriched in prompt leptons (right).}
\label{fig:lep_mva}
\end{figure}

The remaining physics objects include jets, taus, and missing transverse energy (MET). Jets are clustered with the anti-kT algorithm using a cone
size of 0.4. The transverse momenta of jets must be greater than 25 GeV. Finally b-tagging is accomplished with an MVA tagger, where
two working points, loose, and medium are used to identify b-quark jets. Taus are reconstructed and tagged with an MVA discriminator, specifically
tuned for $\tth$ events, with a transverse momentum requirement of greater than 20 GeV. The MET used is the standard MET used across the 
CMS collaboration. 

\section{Event selection}
The event selections are also similar between ATLAS and CMS, and are primarily defined by the tight lepton multiplicity. The
selection with the highest statistics is the two-lepton same-sign selection with no hadronically decaying taus. In addition
to the nominal requirements, this selection requires lepton transverse momenta be greater than 25 GeV and the presence of
at least 4 jets. The next selection is identical to the previous, but requiring exactly
1 hadronic tau. Next are the three-lepton and four-lepton categories. While the ATLAS analysis uses each of these, CMS combines
them into a greater than or equal to three-lepton category. Here, the sum of the three lepton electric charges must be $\pm$1,
there must be greater than or equal to 2 jets and finally a veto on the presence of two same-flavor opposite-sign
leptons within 10 GeV of the Z boson mass is required. Additionally, among the jets counted in the jet multiplicity requirement, there must be
at least 2 jets passing the loose working point, or at least one jet passing the medium working point of the CMS b-tagging discriminator. 

\section{Background estimation}
The main backgrounds can be classified as reducible and irreducible. The reducible backgrounds consist primarily of non-prompt leptons from
b-quark decays in semi-leptonic $\ttbar$, but also include prompt leptons with incorrectly measured electric charge. Most importantly,
the reducible backgrounds are estimated via data-driven control regions. The irreducible backgrounds are estimated from Monte Carlo (MC)
samples and consist of $\ttv$ and diboson processes. The irreducible backgrounds are referred to as such because the signal regions do
not explicitly veto events from these processes.

\section{Signal extraction}
The signal extraction strategy relies on a two-dimensional MVA technique. There are two boosted decision tree (BDT) MVAs,
each trained to discriminate against a single background. The backgrounds targeted individually are the two largest backgrounds,
$\ttv$, where V is a vector boson, and the fake lepton background from $\ttbar$. The output from each BDT is plotted on
a separate axis, forming a two-dimensional shape, which is then binned, forming a one-dimensional shape in each category.

Additionally, the categories described previously are split by the sign of the sum of the lepton
charges, and by the presence (or absence) of two medium b-tagged jets, referred to as the b-tight and b-loose categories as described below
in Figure~\ref{fig:categories}. This
additional categorization provides additonal discrimination power. The BDTs are also trained separately in two-lepton same-sign, and 
three-lepton categories. The inputs for these BDTs consist of lepton and jet kinematics, solid angles, and MET variables. In addition
to these variables, the outputs from a Matrix Element Method (MEM) are also used as inputs to the BDT trained against the
$\ttv$ background (for the three-lepton category only). The MEM output is calculated using a signal hypothesis of $\tth$,$H\rightarrow$WW
assuming three final state leptons. The background hypothesis is $\ttv$. 

\begin{figure}[htb]
\centering
\includegraphics[height=2in]{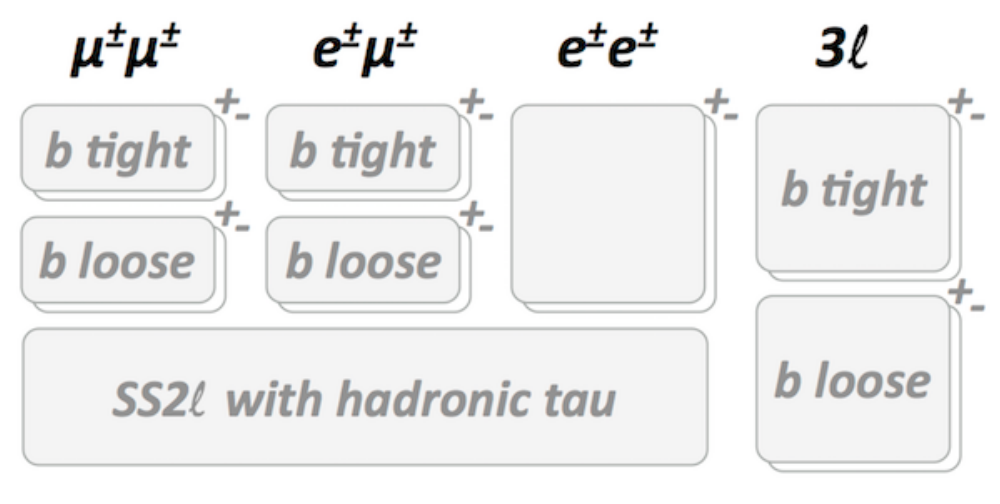}
\caption{A summary of the categories used for the signal extraction.}
\label{fig:categories}
\end{figure}

\section{Results}
The results were produced with the 2015 and 2016 13 TeV LHC Run II datasets and are summarized below in Figure~\ref{fig:tth_results}.
In total, 15.2 $fb^{-1}$ of data was analyzed. Using 12.9 $fb^{-1}$ of data collected in 2016 alone, the analysis measured
95$\%$ C.L. upper limits of 4.6 ($1.7^{+0.9}_{-0.5}$) observed (expected), with best-fit signal strength $\mu = 2.7^{+1.1}_{-1.0}$ in the $2lss$ category,
3.7 ($2.3^{+1.2}_{-0.7}$) observed (expected), with $\mu = 1.3^{+1.2}_{-1.0}$ in the $\ge3l$ category,
and 3.9 ($1.4^{+0.7}_{-0.4}$) observed (expected), with $\mu = 2.3^{+0.9}_{-0.8}$ for 2016 combined.
Combining the data collectd in 2015 with that of 2016, the analysis set 95$\%$ C.L. upper limits on $\tth$ production of
3.4 ($1.3^{+0.6}_{-0.4}$) observed (expected), and measured the best-fit signal strength $\mu = 2.0^{+0.8}_{-0.7}$. This corresponds to a significance
of $3.2\sigma$ under the background-only hypothesis.
The leading systematic uncertainties in each analysis include the non-prompt background estimation, pile-up modeling, luminosity, signal and background modeling,
and jet energy corrections. 

\begin{figure}[htb]
\centering
\includegraphics[height=1.8in]{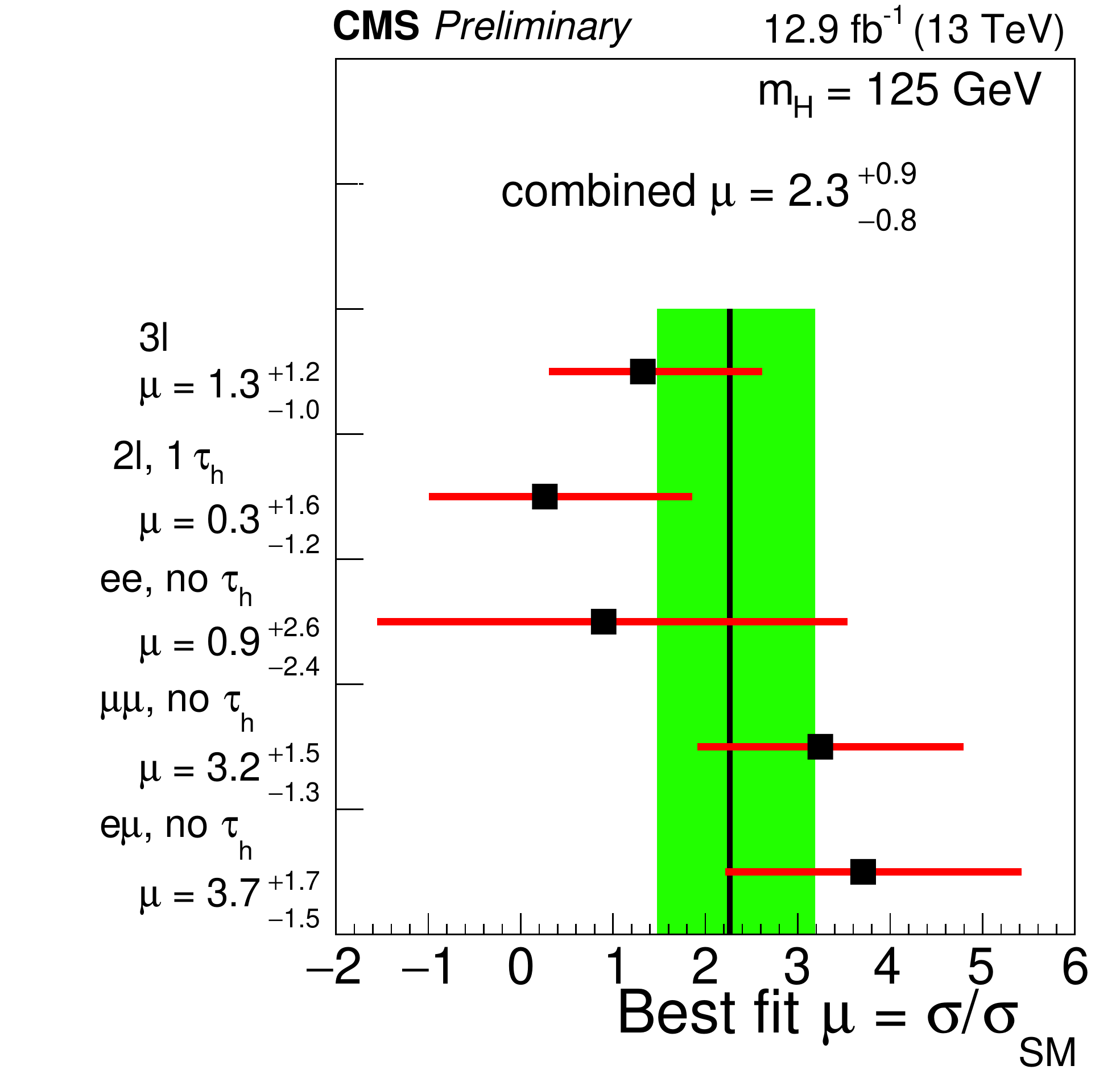}
\includegraphics[height=1.8in]{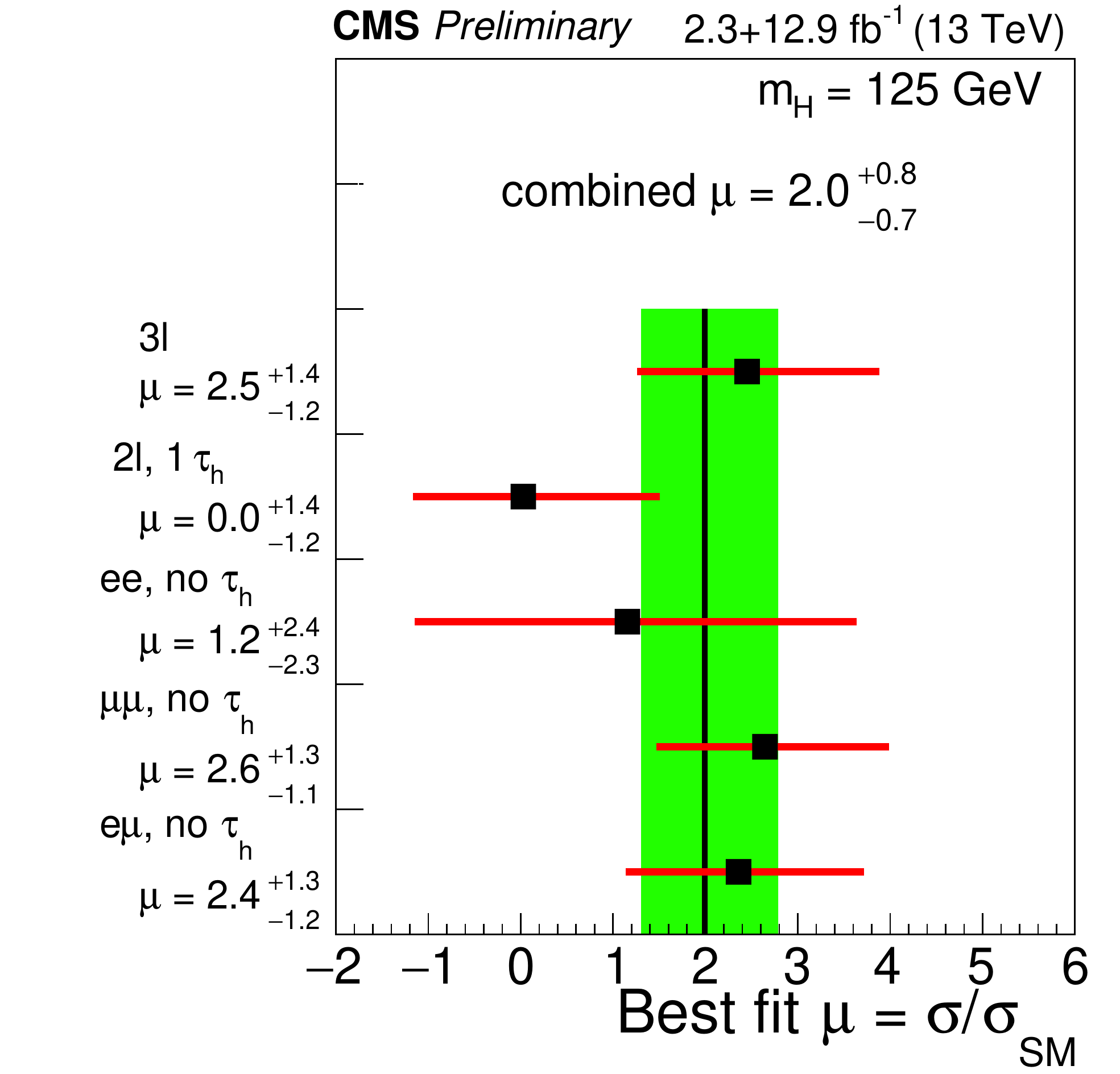}
\includegraphics[height=1.9in]{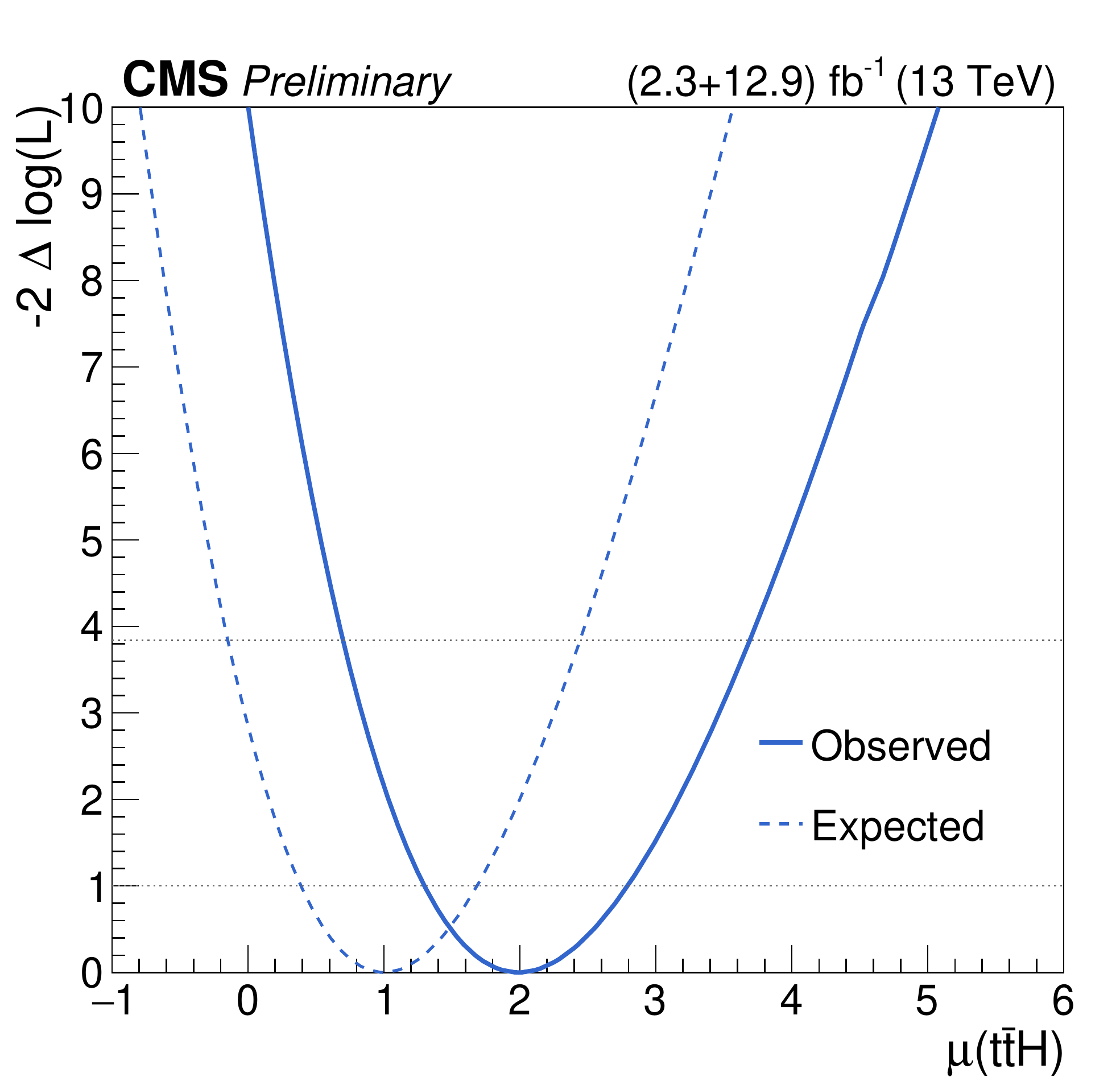}
\caption{A summary of measured upper limits and best-fit $\mu$ with 2016 data only (left), 2015+2016 data combined (middle) and best-fit $\mu$ for 2015+2016 (right)}
\label{fig:tth_results}
\end{figure}

\end{document}